# Perceived Vulnerability to Disease Scale: Factorial structure, reliability, and validity in times of Portugal's COVID-19 pandemic lockdown

Ana Paula Martins

*Department of Mathematics, Faculty of Science, University of Beira Interior, Covilhã, Portugal*

*CMA-UBI Center of Mathematics and Applications of the University of Beira Interior, Covilhã, Portugal*

amartins@ubi.pt
https://orcid.org/0000-0002-3908-821X

María C. Vega-Hernández

*Department of Statistics, University of Salamanca, Salamanca, Spain*

mvegahdz@usal.es

https://orcid.org/0000-0003-4266-4908

Francisca Ribeiro Soares

*Faculty of Health Sciences, University of Beira Interior, Covilhã, Portugal*

francisca.r.soares3@gmail.com
https://orcid.org/0000-0001-5407-7028

Rosa Marina Afonso (Corresponding Author)

*Department of Psychology and Education, University of Beira Interior, Covilhã, Portugal*

*CINTESIS@RISE, Department of Psychology and Education, University of Beira Interior, Covilhã, Portugal*

rmafonso@ubi.pt
https://orcid.org/0000-0003-2111-6873

# Perceived Vulnerability to Disease Scale (PVD): factorial structure, reliability and validity in times of Portugal's COVID-19 pandemic lockdown

The present study examines the factor structure of a Portuguese version of the Perceived Vulnerability to Disease Scale (PVD), designed to assess individual differences in chronic concerns about transmission of infectious diseases. Method: Data from a Portuguese convenience sample (n=1203), collected during the first Covid-19 pandemic lockdown. Results: the scale revealed, through an exploratory factor analysis (EFA) and a confirmatory factor analysis (CFA), a slight superiority of a three-factor model over the existing two-factor models of the 15-item original PVD and of the 10-item PVD established with another Portuguese sample (Ferreira et al., 2022). Conclusions: This higher level of differentiation in terms of a perceived resistance to infectious diseases could be explained by the pandemic context which may have differentiated the responses regarding the perception of Resistance. On the other hand, this new factor increases the comprehensive and evaluative dimension and implications of the construct assessed by PVD.

**Introduction**

Perceived vulnerability to disease are a central construct to the comprehension of social behaviours related to the contraction of infecto-contagious diseases (Duncan et al., 2009; Park et al., 2013). This construct refers by the extent to which individuals perceive themselves to be susceptible to contracting infectious diseases (Bú et al., 2021).

Under the 2019 Coronavirus disease (or COVID-19) which has greatly impacted many of our life certainties, the Perceived vulnerability to disease assumes a central role for the comprehension of the pandemics behaviour and context. Feelings of fear and worry about the future have become more recurrent and our perceived vulnerability to diseases has somehow changed. Safra et al. (2021) point out that the perceived vulnerability to diseases plays a particularly important role to calibrate threat perception. Individuals who perceive themselves as more sensitive to diseases have stronger reactions to disease cues.

The first version of the PVD questionnaire dates to Park et al. (2003) and since then several versions of the questionnaire have been employed. Nevertheless, it was only in 2009 that a version of this questionnaire was published in the scientific literature and psychometrically validated (Duncan et al., 2009). The results revealed two conceptually distinct factors measured by 15 items. One factor (measured by seven items) assesses individuals' beliefs pertaining to their susceptibility to infectious diseases, corresponding to the subscale Perceived Infectability. The second factor (measured by eight items) assesses individuals' discomfort in situations that connote an increased likelihood for the transmission of pathogens, corresponding to the subscale Germ Aversion. Duncan et al., 2009 emphasize that within the domain of social inference, Germ Aversion predicts responses rooted in intuitive appraisals of disease transmission risk, whereas Perceived

Infectability predicts responses informed by more rational appraisals.

To our knowledge, the first Portuguese version of Duncan et al.'s (2009) PVD appeared in Marques (2016). This translated scale was applied to a group of 155 Portuguese citizens in two distinct time periods. A reliability analysis led to a Cronbach's alpha of, respectively, 0.557 and 0.620 for the first and second application of the PVD. From this analysis, two items were removed from the 15-item PVD, conducting to the new alpha values of 0.677 (before) and 0.732 (after). No psychometric validation of the obtained scale was performed and the bifactorial structure found by Duncan et al. (2000) was not considered. Recently, Ferreira et al. (2022) provided a new translation to Portuguese of the 15-item PVD and validated it for the Portuguese population. A convenience sample of 136 participants recruited from the staff and students of three Portuguese universities, was used to determine the factorial structure of the PVD. Exploratory Factor Analysis and Confirmatory Factor Analysis led to a 10-item bifactorial structure for this Portuguese version of the PVD. This 30% reduction of the original 15-item scale extensively narrows the content validity of the Portuguese scale.

Do Bu et al. (2021) adapted and validated the PVD to Brazilian Portuguese. With a sample size of n=201 Brazilians, parallel analysis with 1000 simulation showed appropriateness of retaining three dimensions, which together explained 47,77% of the total variance. Nevertheless, the bifactorial structure of Duncan et al. (2009) was considered, with 13 items retained (items 11 and 12 removed). A similar structure was considered by Stangier et al. (2021) (items 4 and 15 removed) when screening a convenience sample in Germany (n=1358) during lockdown for perceived vulnerability to disease. A Spanish PVD was considered in Díaz et al. (2016), where a sample of n= 878 participants also highlighted the two factors, however, the Germ Aversion factor only showed adequate confirmatory factorial indices and high temporal stability when items

11 and 13 were removed but failed to obtain acceptable internal consistency. Another Spanish version of the scale was provided by Magallares et al. (2017) and applied to 744 university students from all over Spain. A low reliability of reverse items was found and so all six reversed items were removed (items 3, 5, 11, 12 and 13). Confirmatory factor analysis revealed that the items on the short version of the questionnaire (9 items) corresponding to an interrelated two-factor showed the best fit of all the tested models. The Japanese-version PVD (Fukukawa et al., 2014) applied to a sample of n=435 Japanese university students, also has a two-factor structure, but the internal consistency of the Germ Aversion factor was not very high. Another example is the study with Farsi and Arabic Speaking Refugees (Kananian et al., 2022) which worked with the Farsi version and the Arabic version of the PVD. This questionnaire consisting of 28 items (items 4 and 15 removed were removed since they appeared to be no more appropriate to contemporary life conditions) and two subscales (Perceived Infectability and Germ Aversion) was analysed in terms of the total score because the internal consistency of the Germ Aversion subscale obtained was very low and of the other subscale was only moderate.

This study aims to clarify the factor structure of a Portuguese version PVD in a situation like the one experienced with the first closure of the Covid-19 pandemic.

**Materials and methods**

*Participants and procedure*

There were a total of 1203 Portuguese residents survey respondents that fully completed the questionnaire. The respondents ages ranged from 18 to 93 years, with a mean age of 41.7 years (s.d. 17.47 years). Women were the majority among respondents

(n= 818; 68%) and participants lived mainly in the north (n=535, 42,1%) and center (n=421, 35%) regions of Portugal. More detailed sociodemographic information of the participants can be found in Table 1.

**Table 1**
Sociodemographic characteristics of study sample (N=1203)

|  | Frequency | % |
|---|---|---|
| **Gender** |  |  |
| Female | 818 | 68.0 |
| Male | 385 | 32.0 |
| **Age** |  |  |
| 18-25 | 314 | 26.1 |
| 25-34 | 169 | 14.0 |
| 35-44 | 191 | 15.9 |
| 45-54 | 255 | 21.2 |
| 55-64 | 107 | 8.9 |
| ≥65 | 167 | 13.8 |
| **Marital Status** |  |  |
| Single | 507 | 42.1 |
| Married /Partnered | 594 | 74.3 |
| Divorced | 81 | 6.7 |
| Widowed | 21 | 1.7 |
| **Residence Region** |  |  |
| North | 535 | 44.5 |
| Center | 421 | 35.0 |
| Lisbon and *Vale do Tejo* | 176 | 14.6 |
| *Alentejo* | 14 | 1.2 |
| *Algarve* | 23 | 1.9 |
| Archipelagos' *Madeira* and *Açores* | 23 | 1.9 |
| Other | 11 | 0.9 |
| **Degrees** |  |  |
| 1-9 years | 60 | 5.0 |
| Secondary Education (10-12 years) | 301 | 25.0 |
| Higher Education Bachelor´s Degree | 582 | 48.4 |
| Master´s or Doctoral Degree | 260 | 21.6 |
| **Profession** |  | 45.6 |
| Employed | 548 |  |
| Self-employed | 79 | 6.6 |
| Independent worker | 48 | 4.0 |
| Unemployed | 45 | 3.7 |
| Student | 281 | 23.4 |
| Working student | 26 | 2.2 |
| Retired | 169 | 14.0 |
| Not working nor searching (homemaker) | 7 | 0.6 |

The questionnaire was applied online, via e-mail (to personal and professional networks) and posts on social media like Facebook, during the first Portuguese Covid-19 lockdown, from 29th April until 28th May 2020. To avoid missing data, *all responses were mandatory*.

*Instruments*

A questionnaire with socio-demographic questions (e.g., sex, age and residence region), a Portuguese version of Duncan et al.'s (2009) Perceived Vulnerability to Disease Scale (PVD), developed by Marques (2016), and questions concerning COVID-19 was elaborated. This study will focus on the Portuguese PVD. This fifteen-item scale was used to assess participant's perceived vulnerability to an infectious disease, more precisely, perceived infectability (seven-item sub-scale) and germ aversion (eight-item sub-scale). Participants respond to each item on a seven-point scale (from "strongly disagree" to "strongly agree") with approximately half of the items reverse-scored (Duncan et al., 2009). A higher score indicates a severe form of perceived infectability, germ aversion, or perceived vulnerability to disease. Example items include "In general, I am very susceptible to colds, flu and other infectious diseases" (perceived infectability) and "I prefer to wash my hands pretty soon after shaking someone's hand" (germ aversion).

*Ethics Statement*

Informed consent was obtained electronically prior to the collected data from the participants. All procedures performed were previously approved by the ethics committee of the University of Beira Interior (CE-UBI-Pj-2020-037: ID1972)

*Statistical Analysis*

The factor structure determination of the Portuguese PVD was achieved through Exploratory Factor Analysis (EFA) and Confirmatory Factor Analysis (CFA). To avoid optimistic model fit indices and parameter estimates, EFA and CFA were not performed on the same data as suggested by Fokkema and Greiff (2017), Knekta et al. (2019), among others. The original sample (n=1203) was divided into two subsamples with the Solomon method (Lorenzo-Seva., 2021), which is well adapted to the context of factor analysis, performs better than the usual random split approach and provides equivalent subsamples in the sense of the Kaiser-Meyer-Olkin criteria (KMO). An EFA was first performed on half of the sample (n=602) to determine the dimensionality of the PVD and detect possible problematic items. CFA was then performed on the other half of the sample (n=601). The ordinal and nonnormal nature of the data was taken into account by considering polychoric correlations (Holgado–Tello et al., 2010) with unweighted least squares (ULS) estimators in EFA (Forero et al., 2009) and with weighted least square mean and variance adjusted (WLSMV) estimators in the CFA (Yang-Wallentin & Jöreskog, 2010).

Horn's parallel analysis with polychoric correlation (Garrido et al., 2013, Lim & Jahng, 2019) and the Hull method (Lorenzo-Seva et al., 2011) were employed to determine in the EFA the number of factors to retain. Items with factor loadings bigger than 0.5 and no complex or cross-loadings were also retained. The extracted factor solution was finally examined via CFA on the second sample (n=601). Polychoric correlations with WLSMV estimators were here considered (Yang-Wallentin & Jöreskog, 2010 and Li, 2016).

A total of six possible models were tested and compared with the several fit indices. The Chi-square ($\chi 2$) method, the most used to evaluate goodness-of-fit, is highly sensitive to sample size, for models with 75 to 200 cases chi-square is a reasonable measure of fit, but for 400 cases or more it is nearly almost always significant. The usual procedure is to consider this index in association with others (Brown, 2015): Comparative Fit Index (CFI), Tucker-Lewis Index (TLI), Root Mean Square Error of Approximation (RMSEA) in association with 90% Confidence Interval (90% CI), Standardized Root Mean Square Residual (SRMR). RMSEA values less than 0.08 suggest adequate model fit (i.e., a "reasonable error of approximation," RMSEA values less than 0.05 suggest good model fit and models with RMSEA $\geq$ 0.1 should be rejected). Additional support for the fit of the solution would be evidenced by a 90% confidence interval of the RMSEA whose upper limit is below these cut-off values (e.g., 0.08). SRMR values close to 0.08 or below indicate a reasonably good fit. CFI and TLI values in the range of 0.90 and 0.95 may be indicative of acceptable model.

Reliability analysis was obtained in the final 12 item PVD, using Cronbach's alpha for internal consistency. It is the most widely used method for estimating internal consistency reliability.

All statistical analyses were performed using IBM SPSS Statistics software (version 28.0, IBM SPSS) and R version 4.2.1 (R Project for Statistical Computing, www.r-project.org).

**Results**

*Scale items responses*

The analysis of the item boxplots identified 11 outliers and 6 extreme outliers in two items: "Item 5. My past experiences make me believe I am not likely to get sick even when my friends are sick" (4 outliers) and "Item 1. It really bothers me when people sneeze without covering their mouths" (7 outliers and 6 extreme outliers). On the other hand, eighteen cases with high Mahalanobis distance (p<0.001) were identified as potential multivariate outliers. All univariate and multivariate potential outliers were inspected, but no justification was found for removing any of the responses. Mean values for the items ranged from 2.17 to 6.34. Most items had a skewness and kurtosis below |1.00| and only one item (Item 1) had a skewness and kurtosis above |2.00| (Table 2).

**Table 2**
Descriptive statistics of sample item responses

| Items | Mean (sd) | Median | Skewness | Kurtosis |
|---|---|---|---|---|
| **Item 1.** It really bothers me when people sneeze without covering their mouths. | 6.34(1.21) | 7 | -2.36 | 5.82 |
| **Item 2.** If an illness is 'going around', I will get it. | 2.59(1.32) | 2 | 0.68 | 0.02 |
| **Item 3.** I am comfortable sharing a water bottle with a friend. (R) | 2.82(2.04) | 2 | 0.75 | -0.84 |
| **Item 4.** I do not like to write with a pencil someone else has obviously chewed on. | 5.23(2.08) | 6 | -0.79 | -0.84 |
| **Item 5.** My past experiences make me believe I am not likely to get sick even when my friends are sick. (R) | 3.47(1.66) | 4 | 0.11 | -0.91 |
| **Item 6.** I have a history of susceptibility to infectious disease. | 2.17(1.40) | 2 | 1.28 | 1.10 |
| **Item 7.** I prefer to wash my hands pretty soon after shaking someone's hand. | 4.19(1.98) | 4 | -0.02 | -1.23 |
| **Item 8.** In general, I am very susceptible to colds, flu and other infectious diseases. | 3.04(1.63) | 3 | 0.62 | -0.50 |
| **Item 9.** I dislike wearing used clothes because you do not know what the last person who wore it was like. | 3.49(2.23) | 3 | 0.40 | -1.31 |
| **Item 10.** I am more likely than the people around me to catch an infectious disease. | 2.95(1.55) | 3 | 0.59 | -0.35 |
| **Item 11.** My hands do not feel dirty after touching money. (R) | 2.95(1.92) | 2 | 0.65 | -0.83 |
| **Item 12.** I am unlikely to catch a cold, flu or other illness, even if it is 'going around'. (R) | 3.18(1.63) | 3 | 0.33 | -0.72 |
| **Item 13.** It does not make me anxious to be around sick people. (R) | 4.03(1.81) | 4 | -0.1 | -1.00 |
| **Item 14.** My immune system protects me from most illnesses that other people get. (R) | 4.53(1.51) | 5 | -0.52 | -1.21 |
| **Item 15.** I avoid using public telephones because of the risk that I may catch something from the previous user. | 4.01(2.19) | 4 | 0.05 | -1.42 |

*sd indicates standard deviation and (R) indicates reverse scored.*

*Factor analysis*

The division of the sample with Solomon's method originated a first subsample (n=602) with a KMO value of 0.797 (Bartlett's Test of Sphericity $\chi^2(105) = 1821.16$, p < 0.001) and a second subsample (n=601) with a KMO value of 0.790 (Bartlett's Test of Sphericity $\chi^2(105) = 1775.4$, p < 0.001). Both KMO values are very similar to the KMO value of 0.806 (Bartlett's Test of Sphericity $\chi^2(105) = 3506.54$, p < 0.001) from the original sample (n=1203) and show good subsampling factor analysis adequacy. Inter-item corelations (polychoric correlations) for both subsamples (Figure 1), reveal that no item had unacceptably inter-item relationships (at least one $|\rho|>0.3$).

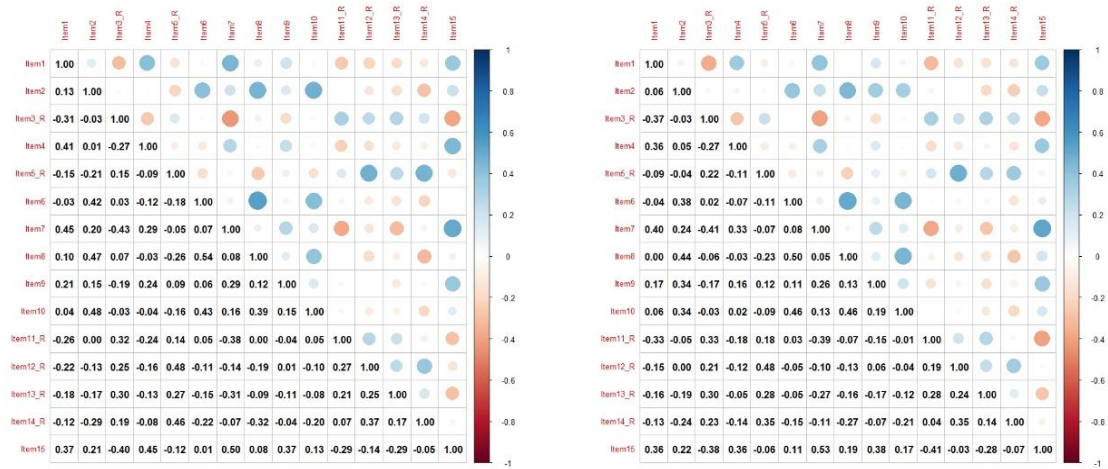

**Figure 1** Inter-item correlations (polychoric correlations) for the two subsamples (n=602 (left) and n=601 (right)). _R indicates a reverse-scored item.

EFA was then performed on the first subsample and CFA on the second, with the considerations of polychoric correlations (Holgado–Tello et al., 2010) in both analyses. To determine the number of factors to retain a first graphical strategy was used (Figure 2). Horn's parallel analysis with unweighted least-squares estimator (ULS) showed that four factors should be retained (Figure 2). Lim & Jahng (2019) point out that when performing a parallel analysis, the number of factors within ±1 range of the estimate can be considered as viable candidates, which agrees with Hull's method (Lorenzo-Seva et al., 2011), with CAF, CFI and RMSEA indices, that support a three-factor retention.

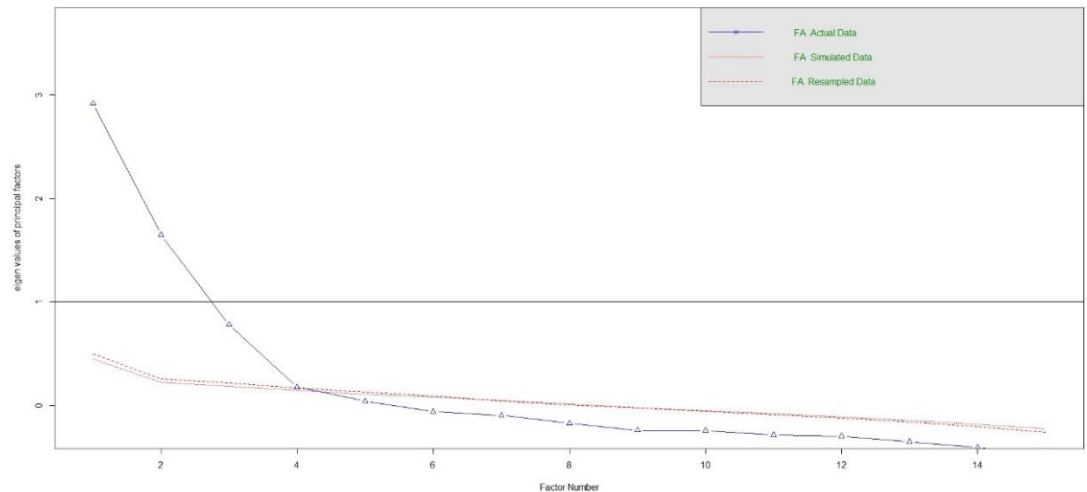

*Figure 2* Parallel analysis, with polychoric correlations and ULS estimator, scree plot.

A three-factor solution, with promax rotation and unweighted least square estimation (ULS), was forced. This solution accounted for 37% of the variance, with factor 1 explaining 17%, factor 2 explaining 13% and factor 3 explaining 10% of the variance. The inter-factor correlations were 0.11 for factors 1 and 2, -0.22 for factors 1 and 3 and 0.25 for factors 2 and 3. The first factor replicated the Germ Aversion (GA) factor from the original scale. The second factor included four items that access personal susceptibility to infectious diseases and so it was designated by Perceived Infectability (PI) as in the original scale. The novel factor, factor 3, included three items that access beliefs about immunological functioning and therefor it was denoted by Perceived Resistance (PR). Table 3 lists each of the 15 items along with factor loadings on each factor and communality, uniqueness and complexity scores for each item.

**Table 3**
Factor loadings of 15 items on the perceived vulnerability to disease scale adapted to Portuguese language for COVID-19 pandemic situation.

| | Factor1(GA) | Factor 2 (PI) | Factor 3 PR) | Communality | Uniqueness | Complexity |
|---|---|---|---|---|---|---|
| **Item 15.** I avoid using public telephones because of the risk that I may catch something from the previous user. | **0.75** | 0.09 | 0.11 | 0.55 | 0.45 | 1.1 |
| **Item 7.** I prefer to wash my hands pretty soon after shaking someone's hand. | **0.73** | 0.11 | 0.10 | 0.53 | 0.47 | 1.1 |
| **Item 1.** It really bothers me when people sneeze without covering their mouths. | **0.56** | -0.01 | -0.07 | 0.34 | 0.66 | 1.0 |
| **Item 4.** I do not like to write with a pencil someone else has obviously chewed on. | **0.55** | - 0.13 | - 0.01 | 0.31 | 0.69 | 1.1 |
| **Item 3.** I am comfortable sharing a water bottle with a friend. (R) | **-0.54** | 0.12 | 0.16 | 0.37 | 0.63 | 1.3 |
| **Item 9.** I dislike wearing used clothes because you do not know what the last person who wore it was like. | **0.45** | 0.20 | 0.26 | 0.25 | 0.75 | 2.0 |
| **Item 11.** My hands do not feel dirty after touching money. (R) | **-0.42** | 0.16 | 0.19 | 0.26 | 0.74 | 1.8 |
| **Item 13.** It does not make me anxious to be around sick people. (R) | **- 0.31** | - 0.07 | 0.23 | 0.21 | 0.79 | 2.0 |
| **Item 8.** In general, I am very susceptible to colds, flu and other infectious diseases. | - 0.07 | **0.68** | - 0.17 | 0.52 | 0.48 | 1.2 |
| **Item 6.** I have a history of susceptibility to infectious disease. | - 0.12 | **0.67** | -0 .10 | 0.47 | 0.53 | 1.1 |
| **Item 2.** If an illness is 'going around', I will get it. | 0.10 | **0.66** | -0.05 | 0.48 | 0.52 | 1.1 |
| **Item 10.** I am more likely than the people around me to catch an infectious disease. | 0.04 | **0.63** | 0.00 | 0.40 | 0.60 | 1.0 |
| **Item 5.** My past experiences make me believe I am not likely to get sick even when my friends are sick. (R) | 0.07 | -0.10 | **0.74** | 0.56 | 0.44 | 1.1 |
| **Item 12.** I am unlikely to catch a cold, flu or other illness, even if it is 'going around'. (R) | - 0.10 | 0.01 | **0.64** | 0.46 | 0.54 | 1.0 |
| **Item 14.** My immune system protects me from most illnesses that other people get. (R) | 0.04 | -0.23 | **0.54** | 0.37 | 0.63 | 1.4 |

*Items corresponding to each of the three factors (subscales) are listed according to the strength of their factor loading. (R) indicates reverse-scored.*

Items 9, 11(R) and 13(R) registered low loadings (<0.5), small communality values, high complexity and cross-loadings for the sample size. The decision of removing these items was confirmed by several factor analyses with different oblique rotations.

The 12-items trifactorial solution, extracted from the EFA, was confirmed via CFA on the second sample, with WLSMV estimator (Model 5.1 in Table 4). Results revealed an acceptable global adjustment $\chi^2(51)=224.320$, CFI=0.922, TLI=0.899, RMSEA=0.075 (90% CI [0.065, 0.085]) and SRMR=0.062. In this model, all items reached high factor weights (Figure 3). This model was compared with other alternative models (Table 4). Together with the Duncan et al.'s (2009) original bifactorial structure with the 15 items of the scale (Model 1), the models proposed by Do Bú et al. (2021) (Model 2: bifactorial structure without items 11_R and 12) and by Ferreira et al. (2022) (Model 3: bifactorial structure without items 2, 5_R, 6, 9 and 13R) were also considered due to cultural and linguistic similarities, since both consider Portuguese translations of the scale and Ferreira et al.'s (2022) also focuses on the Portuguese population. Stangier et al.'s (2021) bifactorial model with 13 items (Model 4: without items 4 and 15) was also considered because of the identical data collection period. Finally, two trifactorial models were evaluated, the one obtained from the EFA (Model 5.1 – without items 9, 11_R and 13_R) and the one with all the 15 items of the scale (Model 5.2). All models were tested with our second subsample (n=601) and are presented in Table 4.

**Table 4**
Goodness of fit statistics for confirmatory factor analysis models of the Portuguese PVD (n=601).

|  | Model 1 | Model 2 | Model 3 |
|---|---|---|---|
|  | Two factors PI & GA 15 items | Two factors PI & GA 13 items | Two factors: PI & GA 10 items |
| $\chi^2$(df) | 776.788(89)* | 423.701(64)* | 188.631(34)* |
| CFI | 0.757 | 0.842 | 0.897 |
| TLI | 0.713 | 0.807 | 0.864 |
| RMSEA (90% CI) | 0.113 (0.106, 0.121) | 0.097 (0.088, 0.106) | 0.087 (0.075, 0.099) |
| SRMR | 0.094 | 0.081 | 0.066 |
| Loadings range | PI: 0.43 to 0.70 | PI: 0.30 to 0.73 | PI: 0.41 to 0.60 |
|  | GA: 0.41 to 0.74 | GA: 0.42 to 0.75 | GA: 0.47 to 0.72 |
| Item loadings <0.5 | PI: 5_R; 12_R | PI: 14_R; 5_R | PI: 12_R |
|  | GA: 4; 9; 13_R | GA: 4; 9; 13_R | GA: 4 |

|  | **Model 4** | **Model 5** | **Model 6** |
|---|---|---|---|
|  | Two factors | Three factors | Three factors |
|  | PI & GA | PI & GA & PR | PI & GA & PR |
|  | 13 items | 12 items | 15 items |
| $\chi^2$(df) | 2305.748(78)* | 224.320(51)* | 5.33* |
| CFI | 0.709 | 0.922 | 0.867 |
| TLI | 0.645 | 0.899 | 0.839 |
| RMSEA (90% CI) | 0.130 (0.121, 0.139) | 0.075 (0.065, 0.085) | 0.085 (0.077, 0.093) |
| SRMR | 0.101 | 0.062 | 0.074 |
| Loadings range | PI: 0.44 to 0.70 | PI: 0.58 to 0.77 | PI: 0.6 to 0.76 |
|  | GA: 0.37 to 0.66 | GA: 0.49 to 0.71 | GA: 0.39 to 0,73 |
|  |  | PR: 0.62 to 0.67 | PR: 0.61 to 0.67 |
| Item loadings <0.5 | PI: 5_R; 12_R; | GA: 4 | GA: 4; 9; 13_R |
|  | GA: 9 |  |  |

_R denotes reverse-scored. Loadings are reported in absolute value. *p<0.01. PI – Perceived Infectability; GA – Germ Aversion; PR – Perceived Resistance. Model 1- Duncan et al. (2009); Model 2 – without items 11_R and 12_R Do Bú et al. (2021); Model 3 – without items 2, 5_R, 6, 9 and 13R Ferreira et al. (2022); Model 4 – without items 4 and 15 Stangier et al. (2021); Model 5 – without item 9, 11_R and 13_R.

Models 1, 2, 3 and 4 obtained inadequate global and local adjustment values. On the other hand, Models 5 and 6 indicate a reasonably good fit. Therefore, the goodness-to-fit adjustment indices indicate a tridimensional structure of the Portuguese PVD. Considering the better fit of the trifactorial structure obtained from the EFA (Model 5.1), this model was adopted.

*Internal consistency*

Internal consistency for all three factors of the final model obtained was acceptable, with Cronbach's alpha values for GA=0.67 (95% CI [0.63, 0.71]), PI=0.71 (95% CI [0.67, 0.74]) and PR= 0.63 (95% CI [0.58, 0.68]). The lower PR value is due to the fact that only three items form this factor (Figure 3).

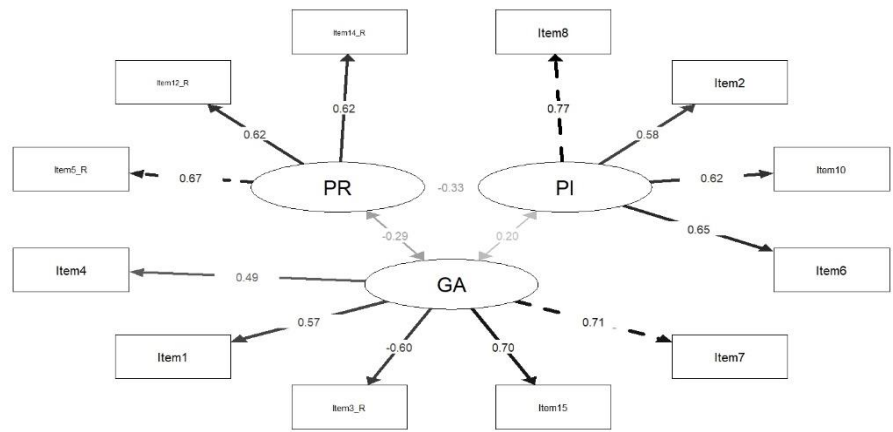

**Figure 3** Path diagram of the trifactorial structure (12 items) of the Portuguese PVD (PI – Perceived Infectability; GA – Germ Aversion; PR – Perceived Resistance). _R indicates a reverse-scored item.

**Discussion and conclusions**

This study uses a wide sample and robustly clarifies the factor structure of the Portuguese PVD version. The recently developed Solomon method was employed to split the original sample into two subsamples, avoiding in this way optimistic model fit indices and parameter estimates when performing the factor analysis, which constitutes a novelty in this type of studies. Nevertheless, the very specific social and sanitary context of the COVID-19 pandemic, in which the study was developed might have had implications on the participants' answering behaviour and must be considered when using and interpreting the proposed version of the questionnaire.

The statistical analysis performed revealed a three-factor structure, with three removed items, for the Portuguese PVD version as opposed to the existing two factor structure. The removed items are: "I dislike wearing used clothes because you do not know what the last person who wore it was like"; "My hands do not feel dirty after touching Money" and "It does not make me anxious to be around sick people", which have low loadings, small communality values, high complexity, and cross-loadings for the sample size. These factors assess the GA scale factor, the one with the larger number of items. Removing those items improves the questionnaire's properties, making it sturdier and smaller, which optimizes its applicability.

Concerning the scale's new 3 factor structure, unlike the known two factor structure from the original PVD scale (Duncan et al., 2009), and in subsequent studies and scale adaptations for different countries (e.g., PVD adaptation for Brazil by Bú et al., 2021), it increases the scale's comprehensive dimension, as well as its use and implications. Although some studies used the two factor PVD subscales (e.g., Murray et al., 2013; Makhanova et al., 2015), other studies did not find satisfactory reliability for these subscales, namely for the Germ Aversion (e.g., Díaz et al., 2016, Miller & Maner, 2012; Wu & Chang, 2012). In this study, the original version's dimensions and various adaptations is maintained: Germ Aversion (GA) and Perceived Infectability (PI). As for the dimension of Germ Aversion (GA), just like the scale's original version, it assesses the aversive affective reactions to situations in which there is a high probability of pathogen transmission. As for the dimension of PI, the original scale assesses beliefs related to immunological functioning and the perceived susceptibility to getting infectious diseases. In the proposed version of this study, this dimension focuses on assessing the perceived susceptibility to getting infectious diseases, with the inclusion of the new factor of Perceived Resistance (PR), which assesses beliefs related to immunological functioning, defences and resistances. Due to this factor reorganization, the scale factors are now assessed by a smaller number of items, while guaranteeing the robustness with which they are assessed. The PI factor is now assessed by 4 items instead of 7, the GA factor is now assessed by 5 items instead of 8, and the new PR factor is assessed by 3 items, which corresponds to the three-dimensional structure of the Portuguese PVD with the goodness-to-fit adjustment (Model 5).

The Perceived Resistance (PR) factor provides a greater specificity of factors and may increase the construct's understanding. To the Germ Aversion (GA) which predicts responses rooted in intuitive appraisals of disease transmission risk and to the Perceived Infectability (PI) which predicts responses informed by more rational appraisals (Duncan et al., 2009), this study adds the Perceived Resistance (PR) concerning protective elements related to vulnerability, which may be assessed and explored in other studies as a predictor of health protecting behaviours. PR is assessed with the following items: "My past experiences make me believe I am not likely to get sick even when my friends are sick"; "I am unlikely to catch a cold, flu or other illness, even if it is 'going around'" and "My immune system protects me from most illnesses that other people get." As such, this factor provides an assessment of the person's perception of their resources to resist the illness and maintain their health. The creation of this new scale factor may be related to the fact that participants answered the questionnaire under a pandemic context, which may have alerted the population for the importance of defences and the functioning of the immune system when dealing with infectious diseases. Although no studies were found about the content of the formal and informal information shared in Portugal on infectious diseases during the 1st COVID-19 pandemic lockdown, this study may alert to the fact that people now differentiate this factor in assessing their vulnerability to the disease. The fact that information was shared about how COVID-19's consequences are more severe among vulnerable people may have highlighted the specificity of perceived resistance that was not differentiated in previous applications of the scale.

In relation to previous Portuguese versions of PVD, both the first Portuguese version by Marques (2016), translated and applied to a group of 155 Portuguese citizens, and more recently the new translated and validated Portuguese version by Ferreira et al. (2022), with a convenience sample of 136 participants recruited among university staff and students, this study includes important updates. On one hand, the study is developed with a larger sample. Even though it's still a convenience sample, having 1203 participants from different geographical areas, different ages and different educational backgrounds increases the sample's representativeness. Moreover, in this study exploratory factor analysis and confirmatory factor analysis were not performed on the same sample, contrarily to the study of Ferreira et al. (2022). On the other hand, since the collection is placed under a specific pandemic context of high vulnerability to infectious diseases, it provides opportunities to explore and appraise the assessment and understanding potential of this scale's construct in such a context. This issue may, however, question whether the new Portuguese version of this scale may be used in a post-pandemic context. The fact that data were collected during the COVID-19 pandemic may have influenced participants' answers, leading them to agree with the items more frequently, as their anxiety and fear of contracting COVID-19 or thoughts of death itself increased during the pandemic (Pakpour & Griffiths, 2020; Silva et al., 2020). Effectively, the current context does not represent the normal life and health parameters. However, it also raises the question of whether the post-pandemic context will be, in terms of perception related to disease vulnerability, closer to the pre-pandemic or pandemic contexts. Taking into consideration the impact of the pandemic in society and behaviours, we may put into question the use of this version of the scale with the presented factor structure, regardless of any considerations on the specificities of the context in which these data were collected.

In conclusion, this study presents important information about PVD factorial structure for the Portuguese population, showing the presence of three dimensions, adding a new factor called Perceived Resistance to the two factors of the original scale and previous studies, Germ Aversion (GA) and Perceived Infectability (PI), which increases the understanding of the construct within the original scale and previous studies. The advantage of obtaining a score for each dimension should be important not only for the individual use of the questionnaire in clinical setting, but also in collective terms of public health, to understand the positioning related to perceived vulnerability and possible dimensions upon which it might be necessary to act in order to promote more adaptative behaviours.

**Disclosure statement**

The authors report there are no competing interests to declare.

**Data availability statement**

The data that support the findings of this study are available from the first author, Ana Paula Martins, upon reasonable request, after approval by the ethics